\documentclass[ 
    aps,
    prl,
    twocolumn,
    letterpaper, 
    10pt,
    superscriptaddress, 
    showpacs,
    showkeys,
    notitlepage,
    amsmath, 
    amssymb, 
    floatfix 
]{revtex4-1}

\usepackage{graphicx}
\usepackage{dcolumn}
\usepackage{bm}
\usepackage{color}
\usepackage{amssymb}
\usepackage{amsmath}
\usepackage{hyperref}
\usepackage{bbold}
\usepackage{accents}

\newcommand{\harpvecsign}{\scriptscriptstyle\text{\tiny$\leftrightarrow$}}
\newcommand{\harpoonvec}[2]{%
  \ifx\displaystyle#1\doalign{$\harpvecsign$}{#1#2}\fi
  \ifx\textstyle#1\doalign{$\harpvecsign$}{#1#2}\fi
  \ifx\scriptstyle#1\doalign{\scalebox{.6}[.9]{$\harpvecsign$}}{#1#2}\fi
  \ifx\scriptscriptstyle#1\doalign{\scalebox{.5}[.8]{$\harpvecsign$}}{#1#2}\fi
}
\newcommand{\doalign}[2]{%
 {\vbox{\offinterlineskip\ialign{\hfil##\hfil\cr#1\cr$#2$\cr}}}%
}

\newcommand{\bel}{\begin{equation}}
\newcommand{\eel}{\end{equation}}

\newcommand{\skyp}[1]{}

\newcommand{\fr}{\frac}

\newcommand{\ee}{\end{equation}}
\newcommand{\be}{\begin{equation}}
\newcommand{\mbf}{\mathbf}

\newcommand{\bal}{\begin{eqnarray} }
\newcommand{\eal}{\end{eqnarray}}
\newcommand{\ba}{\begin{eqnarray*}}
\newcommand{\ea}{\end{eqnarray*}}

\newcommand{\reffig}[1]{Fig.~\ref{#1}}

\newcommand{\ket}[1]{| #1 \rangle}
\newcommand{\bra}[1]{\langle #1 |}

\newcommand{\bq}{{\mathbf q}}
\newcommand{\br}{{\mathbf r}}

\newcommand{\bR}{{\mathbf R}}

\newcommand{\bk}{{\mathbf k}}

\newcommand{\bG}{{\mathbf G}}
\newcommand{\pa}{{\partial}}

\newcommand{\refeq}[1]{Eq.~\eqref{#1}}

\begin{document} 

\title{ Topological Quantum Optics in Two-Dimensional Atomic Arrays}

\author{J. Perczel}
\affiliation{Physics Department, Massachusetts Institute of Technology, Cambridge, MA 02139, USA}
\affiliation{Physics Department, Harvard University, Cambridge,
MA 02138, USA}

\author{J. Borregaard}
\affiliation{Physics Department, Harvard University, Cambridge,
MA 02138, USA}

\author{D. E. Chang}
\affiliation{ICFO - Institut de Ciencies Fotoniques, The Barcelona Institute of Science and Technology, 08860 Castelldefels, Barcelona, Spain}

\author{H. Pichler}
\affiliation{Physics Department, Harvard University, Cambridge,
MA 02138, USA}
\affiliation{ITAMP, Harvard-Smithsonian Center for Astrophysics, Cambridge, MA 02138, USA}

\author{S. F. Yelin}
\affiliation{Physics Department, Harvard University, Cambridge,
MA 02138, USA}
\affiliation{Department of Physics, University of Connecticut, Storrs, Connecticut 06269, USA}

\author{P. Zoller}
\affiliation{Institute for Theoretical Physics, University of Innsbruck, A-6020 Innsbruck, Austria}
\affiliation{Institute for Quantum Optics and Quantum Information of the Austrian Academy of Sciences, A-6020 Innsbruck, Austria}

\author{M. D. Lukin}
\affiliation{Physics Department, Harvard University, Cambridge,
MA 02138, USA}


\date{July 17, 2017}

\bigskip
\bigskip
\bigskip
\begin{abstract} 
We demonstrate that  two-dimensional atomic emitter arrays with subwavelength spacing 
constitute topologically protected quantum optical systems where the photon propagation is robust against large imperfections while losses associated with free space emission are strongly suppressed. Breaking time-reversal symmetry with a magnetic field results in gapped photonic bands with non-trivial Chern 
numbers and topologically protected, long-lived edge states. 
Due to the inherent nonlinearity of constituent emitters, such systems provide a platform for exploring quantum optical analogues of interacting topological systems. 
\end{abstract}

\maketitle
 





Charged particles  in two-dimensional systems exhibit  exotic macroscopic behavior in the presence of magnetic fields and interactions. These include the integer \cite{Klitzing1980}, fractional \cite{Tsui1982} and spin \cite{Konig2007} quantum Hall effects. Such systems support topologically protected edge states \cite{Halperin1982,Laughlin1981} that are robust against defects and disorder.  
There is a significant interest in realizing topologically protected photonic systems. 
Photonic analogues of quantum Hall behavior have been studied in gyromagnetic photonic crystals \cite{Haldane2008,Raghu2008,Wang2008,Liu2012,Wang2009,Yu2008}, helical waveguides \cite{Rechtsman2013b},  two-dimensional lattices of optical resonators \cite{Hafezi2011,Hafezi2013,Fang2012} and in polaritons coupled to optical cavities \cite{Karzig2015}. An outstanding challenge is to realize optical systems which are robust not only to some specific backscattering processes but to {\it all} loss processes, including scattering into unconfined modes and spontaneous emission.  Another challenge is to extend these effects into a nonlinear quantum domain with strong interactions between individual  excitations. These considerations motivate the search for new approaches to topological photonics.  




\begin{figure}[h!]
\centering
\includegraphics[width=0.453 \textwidth]{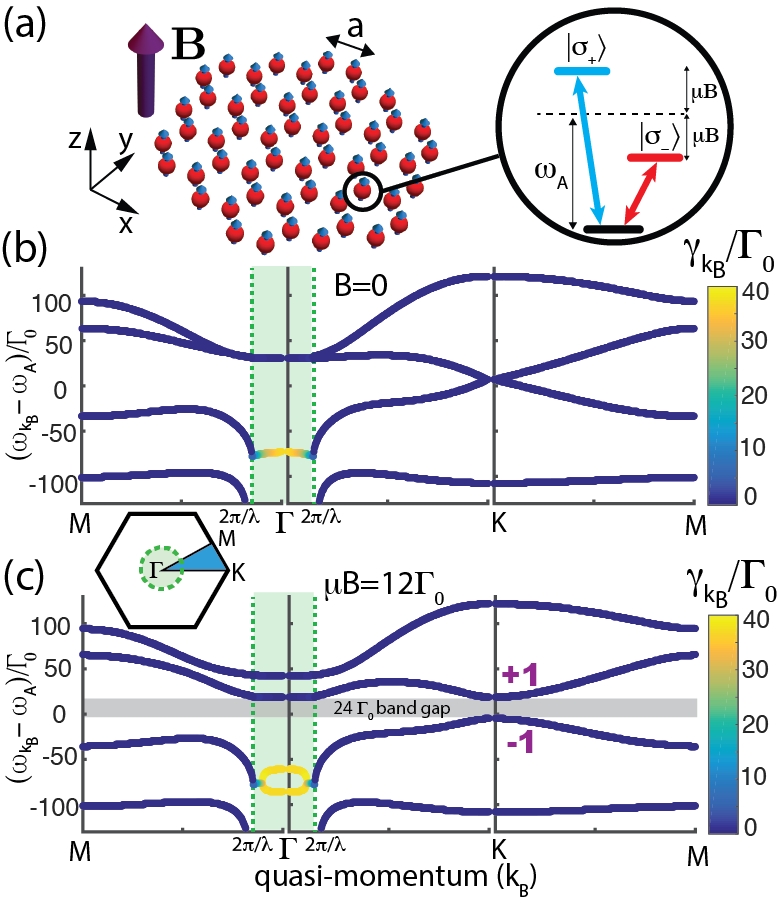}
\caption{
\label{fig:honeycomb}
(a) Honeycomb lattice of atomic emitters with interatomic spacing $a$. Each atom has a V-type level structure with optical transitions to the $\ket{\sigma_+}$ and $\ket{\sigma_-}$ states. A magnetic field breaks the degeneracy via the Zeeman splitting.  (b) Band structure of the lattice with $B=0$. 
Green dashed lines indicate the edges of the free-space light cone. Modes with quasi-momentum $k_\text{B} < \omega_{\bk_B}/c$ couple to free-space modes and are short lived (green shaded region). Decay rates of the modes are color coded. Bands are degenerate at the symmetry points $\mbf K$ and $\mbf \Gamma$. (c) A transverse magnetic field ($\mu B=12\Gamma_0$) opens a gap (grey-shaded region) between topological bands with non-trivial Chern numbers. Relevant parameters are $\lambda=790\text{nm}$, ${\Gamma_0=2\pi\times 6\text{MHz}}$ and $a=0.05\lambda$.} 
\end{figure}

In this Letter, we introduce and analyze a novel  platform for engineering topological states in the optical domain. It is based on atomic or atom-like quantum optical systems \cite{Shahmoon2017}, where 
time-reversal symmetry can be broken by applying magnetic fields and the constituent emitters are inherently nonlinear.    Specifically, we focus on  
optical excitations in  a two-dimensional honeycomb array of closely spaced emitters. We show that such systems maintain topologically protected confined optical modes that are immune to  large imperfections as well as to the most common loss processes such as scattering into free-space modes. Such modes can be used to control individual atom emission, and to create quantum nonlinearity at a single photon level. 

The key idea is illustrated in \reffig{fig:honeycomb}(a).  We envision an  array with interatomic spacing $a$ and quantization axis $\hat z$ perpendicular to the plane of the atoms.
Each emitter has  a V-type level structure with transitions from the ground state to the excited states $ \ket{\sigma_+}$ and $\ket{\sigma_-}$, excited by the corresponding polarization of light \footnote{For simplicity we assume that the $\left| z\right>$ state is far detuned from resonance, e.g. due to Stark shift. However, one can include the $\left| z\right>$ state without detuning as long as there are no polarization-mixing perturbations in the system, as radiation from the $\left|\sigma_+\right>$ and $\left|\sigma_-\right>$ states is completely decoupled from the $\left|z\right>$ transition}. 
The hybridized atomic and photonic states result in confined Bloch modes 
with large characteristic quasi-momenta that for dense atomic arrays significantly exceed the momentum of free-space photons. These confined modes are outside of the so-called ``light cone'' and are decoupled from free space resulting in long-lived, sub-radiant states \footnote{Subradiance in periodic atomic lattices was also discussed recently, in the absence of topology, in Ref.~\cite{Asenjo-Garcia2017}.}.
Atomic Zeeman-shifts induced by a magnetic field, create a bandgap in the optical excitation spectrum, and  the Bloch bands acquire non-trivial Chern numbers. 
The resulting system displays all essential features associated with topological robustness. 
Before proceeding, we note that polar molecules coupled via near-field interactions \cite{Yao2012,Peter2015} and excitons in Moir\'e heterojunctions \cite{Wu2017} have  been shown to give rise to chiral excitations in 2D. In contrast, the present analysis includes both near- and far-field effects as well as scattering to free space. 
We also note that the emergence of Weyl excitations has been recently  predicted \cite{Syzranov2016} in 3D lattices of polar particles. 

In the single excitation case, 
following the adiabatic elimination of the photonic modes, the dynamics of the system (no-jump evolution in the master equation \cite{Gardiner2010}) can be described by the following non-Hermitian spin Hamiltonian \cite{Lukin2016,Antezza2009,Shahmoon2017,Bienaime2012,Guerin2016}
\bal\label{Hamiltonian}
H=\hbar\sum\limits_{i=1}^{N}\sum\limits_{\alpha=\sigma_+,\sigma_-}\left(\omega_A+\text{sgn}(\alpha_i)\mu B-\text{i}\fr{\Gamma_0}{2}\right)\ket{\alpha_i}\bra{\alpha_i}\nonumber \\
+\fr{3\pi \hbar\Gamma_0c}{\omega_A}\sum\limits_{i\neq j}\sum\limits_{\alpha,\beta=\sigma_+,\sigma_-}G_{\alpha\beta}(\br_i-\br_j)\ket{\alpha_i}\bra{\beta_j},\quad
\eal
where $N$ is the number of atoms, $\omega_A=2\pi c/\lambda$ is the atomic transition frequency with wavelength $\lambda$, $\mu B$ is the Zeeman-shift of the atoms with magnetic moment $\mu$ due to an out-of-plane magnetic field $\mbf{B}=B\hat z$ with ${\text{sgn}(\sigma_\pm)=\pm}$. Here, $\Gamma_0=d^2\omega_A^3/(3\pi\epsilon_0\hbar c^3)$ is the radiative linewidth of an individual atom in free space, $c$ is the speed of light, $d$ is the transition dipole moment, $G_{\alpha\beta}(\br)$ is the dyadic Green's function in free space describing the dipolar spin-spin interaction \footnote{See Supplemental Material at [URL] for more details, which includes Refs.~\cite{Antezza2009a,Morice1995,Chew1995,Douglas2015,Dung1998,John1990,Mitsch2014}} and $\br_i$ denotes the position of the atoms. Note that the Hamiltonian in \refeq{Hamiltonian} assumes the atoms are pinned to the lattice. The effect of fluctuating atomic positions is discussed in Ref.~\cite{Note3}.


\begin{figure}
\centering
\includegraphics[width=0.475 \textwidth]{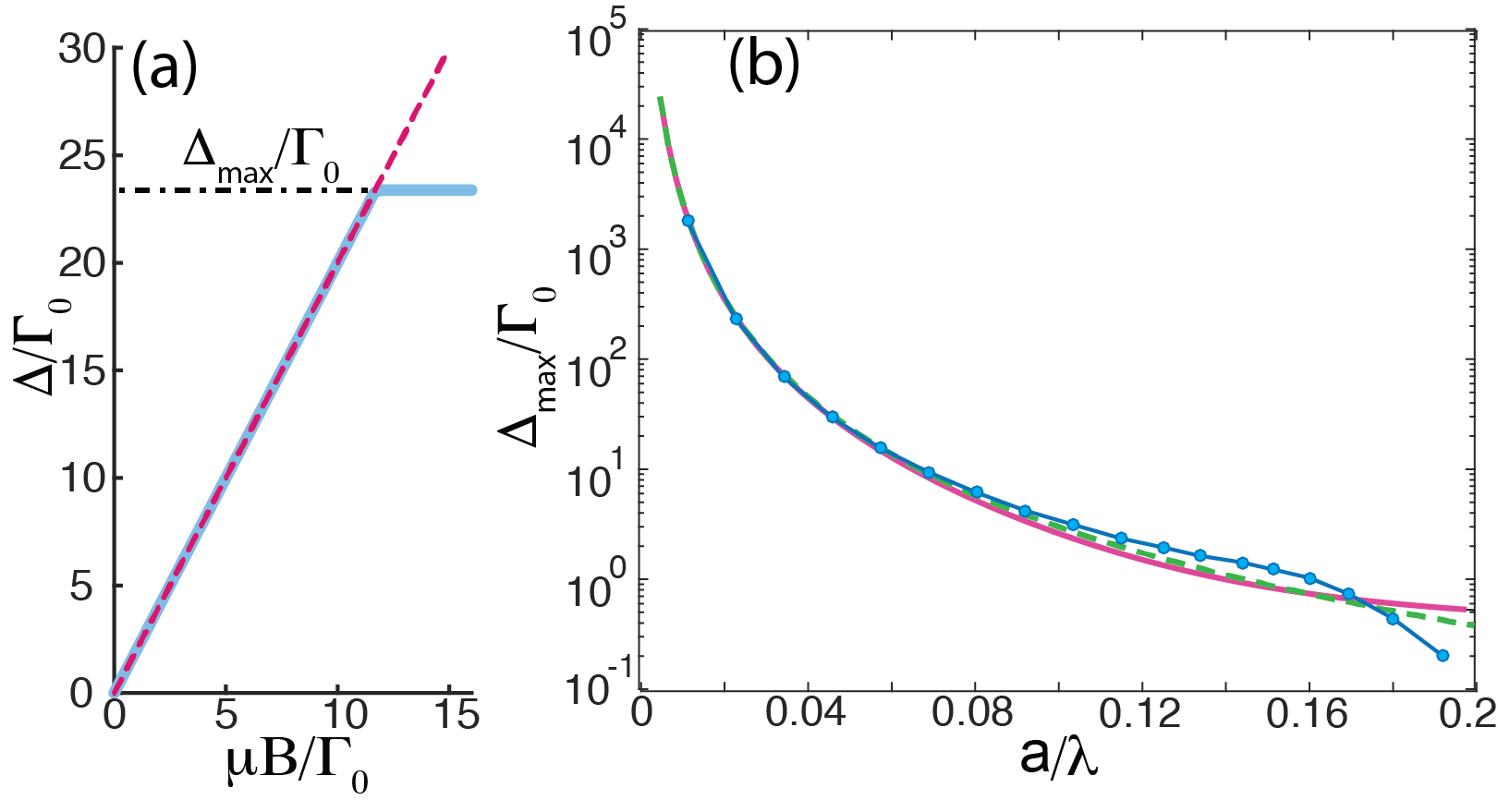}
\caption{
\label{fig:gapsize} (a) Size of the gap between topological bands (blue line) as a function of magnetic field for $a=\lambda/20$. 
(b) 
The maximum gap size $\Delta_\text{max}$ (blue dotted line)
as a function of the interatomic spacing $a$. The solid magenta line shows the dipolar interaction strength $J$ between two atoms with parallel dipole moments. The dashed green line is a phenomenological $J\sim 1/r^3$ fit. For $a\ll \lambda$, $\Delta_\text{max}$ 
scales as $1/a^3$.} 
\end{figure}

For an infinite periodic honeycomb lattice, the single excitation eigenmodes of \refeq{Hamiltonian} are Bloch modes \cite{Bena2009} given by
\bal
\ket{\psi_{\bk_B}}=\sum\limits_n \sum\limits_{b=1,2}e^{\text{i}\bk_B\cdot \bR_n}\Big[c_{+,\bk_B}^{b}\ket{\sigma_{+,n}^{b}}+c_{-,\bk_B}^{b}\ket{\sigma_{-,n}^{b}}\Big],\quad\;
\eal
where the summation runs over all lattice vectors $\{\bR_n\}$, $b=1,2$ labels the two atoms within the unit cell and $\bk_B$ is the Bloch wavevector. For each $\bk_B$ there are four eigenvalues of the form ${E_{\bk_B}=\omega_{\bk_B}-\text{i}\gamma_{\bk_B}}$, where the imaginary part corresponds to the overall decay rate of the modes \cite{Note3}.



\reffig{fig:honeycomb}(b) shows the band structure in the absence of a magnetic field along the lines joining the symmetry points $\mbf M$, $\mbf \Gamma$ and $\mbf K$ of the irreducible Brillouin zone (see inset of \reffig{fig:honeycomb}(c)). The decay rates of the modes ($\gamma_{\bk_B}$) are shown using a color code.
Crucially, we find that the decay rate of some modes can be significantly smaller than $\Gamma_0/2$ due to collective interference effects. 
Green dashed lines at $k_B  = 2\pi/\lambda$ mark the edges of the light cone corresponding to free space modes with dispersion $\omega_{\bk_B}=k_Bc$. 
The modes close to the center of the Brillouin zone ($\mbf \Gamma$) have quasi-momenta $k_\text{B}$ less than the maximum momentum of free space photons at the same energy ($k_\text{B} < \omega_{\bk_B}/c$). These modes couple strongly to free-space modes with matching energy and momentum and decay rapidly \cite{Note3}. 
 In contrast, modes with quasi-momenta greater than the momentum of free space photons ($k_\text{B} > \omega_{\bk_B}/c$), are completely decoupled and do not decay into free space due to the momentum mismatch.

\reffig{fig:honeycomb}(b) also shows that 
the photonic bands are degenerate at the symmetry points $\mbf \Gamma$ and $\mbf K$ in the absence of a magnetic field. These degeneracies originate from the degeneracy of the $\ket{\sigma_+}$ and $\ket{\sigma_-}$ states at zero magnetic field. Due to the lattice symmetries, the degeneracy at the $\mbf \Gamma$ point is quadratic \cite{Chong2008}, while a linear Dirac cone is formed at the $\mbf K$ point \cite{Raghu2008}. Applying an out-of-plane magnetic field lifts this degeneracy and an energy gap forms across the Brillouin zone. 

We explore the topological nature of these bands, by calculating the Chern numbers using 
the method described in Ref.~\cite{Fukui2005}. The sum of the Chern numbers above and below the band gap is $+1$ and $-1$, respectively. The origin of these topological bands can be understood intuitively by noting  that at the $\mbf K$ point the modes separated in energy due to Zeeman splitting have, respectively,  $\hat \sigma_+$  and $\hat \sigma_-$ circular polarizations. The opposite chirality of the bands reflects the time-dependent circular rotation of the electric fields associated with the $\hat\sigma_+$ and $\hat\sigma_-$ polarizations in the $x$-$y$ plane. 


The size of the topological gap at the $\mbf K$ point scales linearly with the magnetic field due to the Zeeman splitting $(2\mu B)$ of the $\ket{\sigma_+}$ and $\ket{\sigma_-}$ states (\reffig{fig:gapsize}(a)), but the gap size is eventually limited to a maximum value $\Delta_\text{max}$ due to the level repulsion between the two upper bands at the $\mbf \Gamma$ point. \reffig{fig:gapsize}(b) shows the maximum gap size as a function of the interatomic spacing $a$ (blue dotted line). The strength of the dipolar coupling $J = 3\pi\Gamma_0c/\omega_A\,  G_{xx}(a)$ between two parallel dipoles at a distance $a$ is also shown. 
The close agreement between the two curves shows that the maximum gap size is determined  by the dipolar interaction strength between the atoms. For $a \ll \lambda$ the maximum gap size has the simple scaling $\Delta_\text{max}\sim 1/a^3$.


\begin{figure}
\centering
\includegraphics[width=0.49 \textwidth]{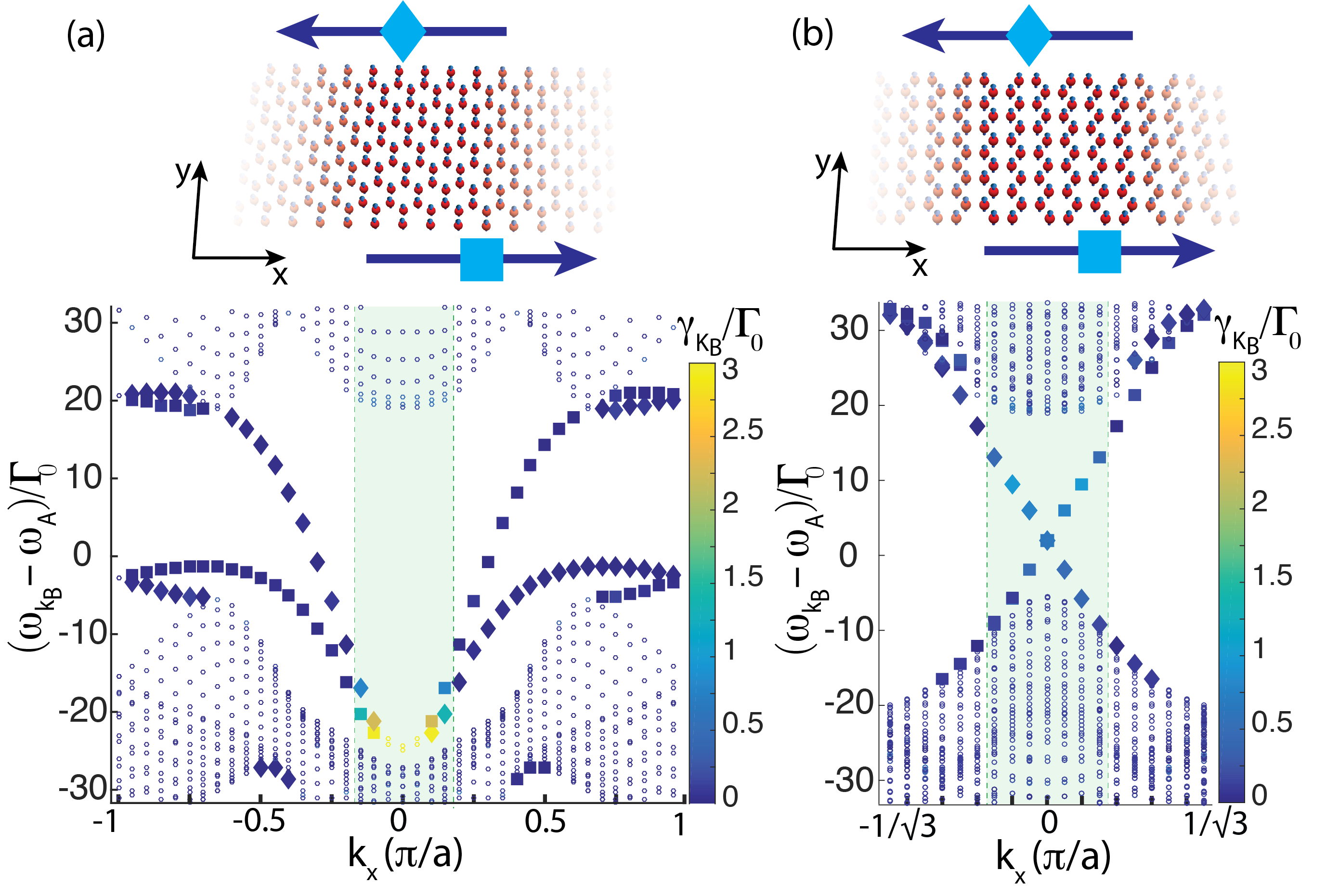}
\caption{
\label{fig:periodic} Topological edge states on the (a) bearded and (b) armchair edges of periodic stripes of atoms. Each edge supports only one unidirectional mode. Modes propagating on the upper (lower) edges of the stripes are marked by diamonds (squares) in the band diagrams. Bulk modes are marked with dots. Decay rates of the modes are color coded. Modes of the bearded (armchair) edges cross the gap with quasi-momentum $k_\text{B} > \omega_{\bk_B}/c$ ($k_\text{B} < \omega_{\bk_B}/c$) making them long (short) lived. Parameters are the same as in \reffig{fig:honeycomb}(c). The spectrum was obtained for the bearded (armchair) edges from an 40x42 (40x41) lattice of atoms with periodic boundary conditions along the first dimension.  States for which the ratio of the total amplitude on the top (bottom) four atom rows to the bottom (top) four rows is greater than 15 are classified as edge states.} 
\end{figure}

\begin{figure}
\centering
\includegraphics[width=0.49 \textwidth]{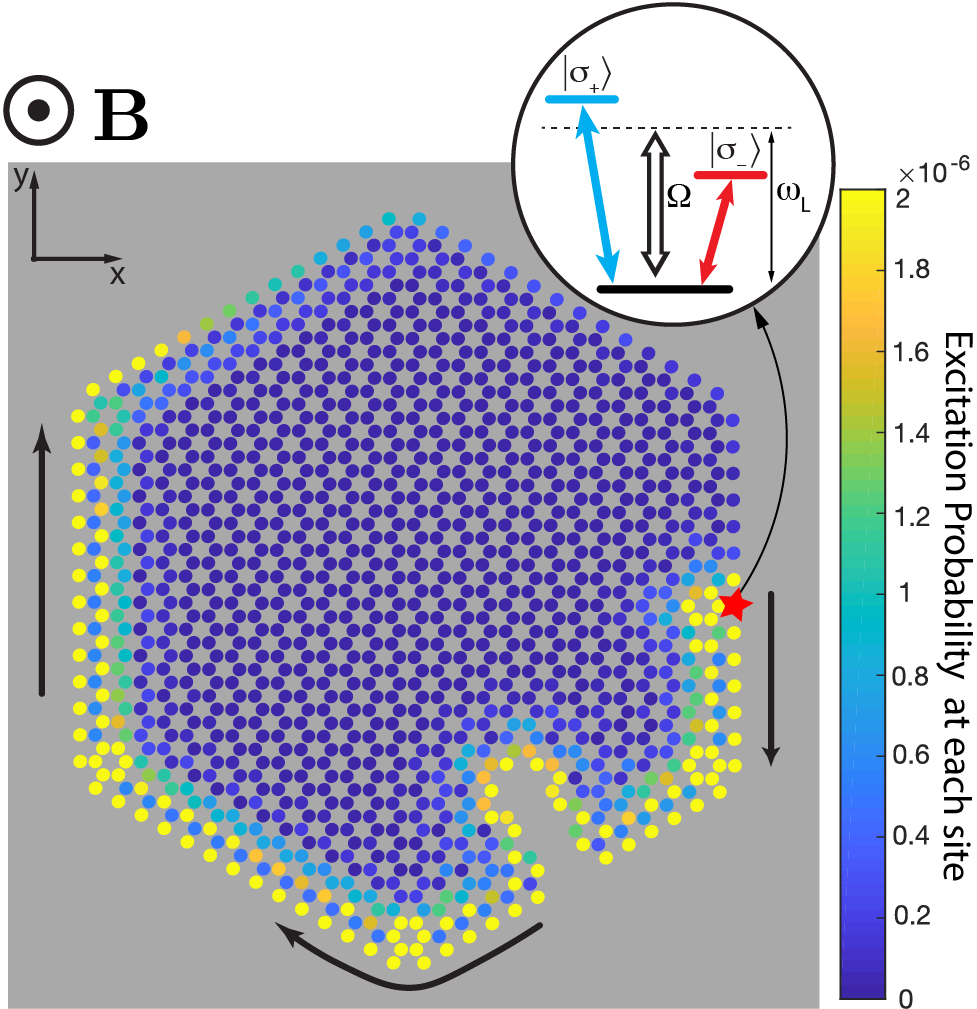}
\caption{
\label{fig:realspace} Snapshot of the time evolution (at $t= 5.7 \Gamma_0^{-1}$) of the system as an atom on the edge (red star) is driven by a laser (inset). The color code shows the excitation probability $|\bra{\psi(t)}\sigma^i_{+}\big>|^2+|\bra{\psi(t)}\sigma^i_-\big>|^2$ at each atomic site $i=1,\dots,N$.
Approximately 96\% of the emitted excitation is coupled into the forward direction and scattering into bulk and backward edge modes is strongly suppressed. The excitation goes around corners and routes around the large lattice defect.
Relevant parameters are $N=1243$, $\lambda=790\text{nm}$, ${\Gamma_0= 2\pi\times6\text{MHz}}$, $a=0.05\lambda$ and $\mu B=12\Gamma_0$. The strength of the drive is ${\Omega=1/5\Gamma_0}$ and the driving frequency is ${\omega_\text{L}=\omega_A+15\Gamma_0}$. The driving laser is adiabatically switched on with a Gaussian profile ${\Omega(t)=\Omega\exp(-[t-1.5\Gamma_0^{-1}]^2/[0.15\Gamma_0^{-2}])}$ for $t<1.5\Gamma_0^{-1}$.} 
\end{figure}

Gaps between topological bands are typically associated with the presence of one-way reflection-free edge modes at the boundaries of a finite system. To explore the spectrum of edge modes in the gap, we calculated the band structure for periodic stripes of atoms in a honeycomb lattice. 
The stripes may have bearded, armchair or zig-zag edges \cite{Plotnik2013,Bernevig2013}. \reffig{fig:periodic} shows the edge geometries and the corresponding band structures of stripes with bearded and armchair edges. Zig-zag edges are discussed in Ref.~\cite{Note3}. 
Edge modes on the lower (upper) edge of the stripe traversing the gap have positive (negative) group velocity and carry energy to the right (left). 
Thus, energy transport by edge modes is unidirectional as a consequence of the broken time-reversal symmetry of the system. If the direction of the magnetic field is flipped, the direction of the energy flow on any given edge is reversed.
Edge modes on bearded boundaries have quasi-momenta $k_\text{B} > \omega_{\bk_B}/c$ while crossing the gap and therefore couple weakly to free-space modes making them long-lived. In contrast, modes on the armchair edges cross the gap with quasi-momenta $k_\text{B} < \omega_{\bk_B}/c$ and the relatively strong coupling to free-space modes makes them short-lived. 
The lifetimes of edge modes are also influenced by the lattice size. Increasing the number of atoms $N$ in a finite lattice, decreases the losses from finite-size effects and increases the lifetimes of long-lived edge modes \cite{Note3}. 

\reffig{fig:realspace} illustrates the unidirectional energy transport. It shows a honeycomb lattice of atoms with an overall hexagonal shape and a large defect on one edge. The geometry was chosen such that in the absence of defects, all boundaries are bearded edges supporting long-lived edge modes. An out-of-plane magnetic field $\mbf B$ induces a band gap of size $\Delta$ in the energy spectrum. An atom on the boundary is adiabatically addressed by a laser at a frequency $\omega_\text{L}$ resonant with the long-lived edge modes in the topmost part of the band gap. 
The laser drives the $\sigma_+$ and $\sigma_-$ transitions of the atom off-resonantly with equal coupling strengths $\Omega$, where $\Omega\ll \Delta$. 
\reffig{fig:realspace} shows a snapshot of the excitation probability of each atom in the lattice. 
Approximately 96\% of the excitation emitted by the driven atom is coupled into the edge modes carrying energy in the forward direction. Coupling into the backward direction or into the bulk modes is suppressed due to topology and the large band gap. These results are qualitatively independent of the relative driving strengths of the $\sigma_-$ and $\sigma_+$ transitions \cite{Note3}. 
The excitation routes around lattice corners with $\sim 97\%$ efficiency and goes around defects of arbitrary shape and size by forming new edge modes at the defect boundaries as shown in \reffig{fig:realspace}, where $\sim 83\%$ of the excitation survives. Atomic emission in the bulk is discussed in Ref.~\cite{Note3}. 


The distance the photon propagates on an edge is set by the ratio of the group velocity and the intrinsic lifetime of the edge modes. The group velocity of the edge modes traversing the gap is $v_g\approx \delta \omega/\delta k_\text{B} \sim \Delta/(\pi/a)$, where $\Delta$ is the size of the energy gap and $a$ is the interatomic spacing. 
Thus for $a\ll \lambda$, 
the maximum group velocity of the edge modes scales as $v_g\sim \Delta_\text{max}/(\pi/a)\sim a^{-2}$. While bearded edges support long-lived modes, any departure from the ideal hexagonal shape of \reffig{fig:realspace} creates a combination of armchair and zig-zag modes that couple more strongly to free-space modes and thus have limited lifetimes. To ensure that only a small fraction of the excitation is lost while the photon is routed around a defect, large group velocities and, therefore, small interatomic spacing is required. 

We note that efficient coupling of individual quantum emitters to a  confined unidirectional channel (Fig.~\ref{fig:realspace})  immediately implies the feasibility of quantum nonlinear interactions between individual photons. This can be understood by considering a `defect atom' placed along the path of the edge excitation. Such an atom can be used to
capture and store an incident photon in a long-lived atomic state, following e.g. Ref.~\cite{Chang2007} (see also Refs.~\cite{Dayan2008,Chen2013,Neumeier2013,Shomroni2014}).  After photon storage, the defect atom  will  form a lattice defect for subsequent incoming photons, which will be routed around this defect and, as a result, will acquire a nonlinear  phase shift.  

Atomic arrays with much smaller interatomic spacing than the transition wavelength ($a\ll \lambda$) could be experimentally realized using state-of-the-art experiments with bosonic Stronium atoms ~\cite{Olmos2013,Syzranov2016}.
Mott insulators in the $^1S_0$ ground state of {$^{84}$Sr} atoms using a 532nm trapping laser have been realized experimentally \cite{Stellmer2014} and the atoms can be further transferred to the metastable
$^3P_0$ state \cite{Akatsuka2008}. Using the long-wavelength $^3P_0$--$^3D_1$ transition with {$\lambda_\text{Sr}=2.6\mu$m} for atom-atom interactions would give $a=2\lambda_\text{laser}/(3\sqrt{3})=\lambda_\text{Sr}/12.7$ in an optical honeycomb lattice. The interatomic spacing could be further reduced to $a=\lambda_\text{Sr}/16.3$ using a 412.8nm `magic wavelength' trapping laser providing equal confinement for the $^3P_0$ and $^3D_1$ states \cite{Olmos2013}. Typical trapping frequencies in Mott insulators are $\sim 5E_\text{recoil}/h$ \cite{Bloch2005}, where $E_\text{recoil}/h\approx 13\text{kHz}$ for Stronium. Since the linewidth is $\Gamma_\text{Sr}=290$kHz for the $^3P_0$--$^3D_1$ transition, the motional states of individual atoms are not well resolved and we expect heating due to photon scattering to be small. The main experimental challenge is to ensure near-unity lattice filling \cite{Bakr2010} and near-uniform excitation of atoms to the $^3P_0$ state. Other approaches to deep subwavelength atomic lattices include utilizing vacuum forces in the proximity of dielectrics ~\cite{Gonzalez-Tudela2015}, using adiabatic potentials \cite{Yi2008}, dynamic modulation of optical lattices \cite{Nascimbene2015} or sub-wavelength positioning of atom-like color defects in diamond nanophotonic devices \cite{Dolde2013,Kolkowitz2015,Sipahigil2016,Iwasaki2015} \footnote{Given the robustness of topological lattices, inhomogeneous broadening present in solid-state systems will not significantly change the results as long as the broadening is small compared to the topological energy gap.}. 


Subwavelength emitter lattices could also be created using monolayer semiconductors, such as transition metal dichalcogenides (TMDCs) \cite{Wang2012,Robert2016,Li2015,Palacios-Berraquero2016,Lin2014,Zhou2017}. Large splitting of the $\sigma_+$, $\sigma_-$ valley polarizations due to interaction-induced paramagnetic responses was recently demonstrated in TMDCs ~\cite{Back2017}. Moir\'e patterns \cite{Hunt2013} could provide deep subwavelength ($a<36$nm) periodic potentials for TMDC excitons and give rise to topological bands and chiral excitonic edge states \cite{Wu2017}. In such Moir\'e heterojunctions the band gaps -- and thus the group velocities of edge states-- are predicted to be small ($\Delta<1\Gamma_0$). However, as our current analysis shows, edge states outside the light cone would be long-lived and thus could still propagate a significant distance along the edges of TMDCs prior to decay into far field modes.

In summary, we have shown that two-dimensional atomic lattices can be used to create robust quantum optical systems featuring band gaps between photonic bands with non-trivial Chern numbers. For a finite lattice, unidirectional reflection-free edges states form on the system boundaries at energies inside the band gap. These edge modes are robust against imperfections in the lattice as well as scattering  and emission into free space. These can be used, e.g. to control emission of individual atoms. We emphasize that, in contrast to linear topological photonic systems, a distinguishing feature of the present approach is the intrinsic, built-in nonlinearity associated with quantum emitters in the lattice, which leads to strong interactions between individual excitations.
Harnessing such interactions could open up exciting possibilities for studying topological phenomena with strongly interacting photons, including quantum optical analogues of fractional Quantum Hall states. These include exotic states, such as those with filling fractions $\nu=5/2$ and $\nu=12/5$, which may feature non-Abelian excitations \cite{Nayak2008}.  In addition, the inherent protection against losses may also be used for the realization of robust quantum nonlinear optical devices for potential applications in quantum information processing and quantum state transfer \footnote{Following the completion of this work, we became aware of the related study R. J. Bettles, J. Min\'{a}\v{r}, I. Lesanovsky, C. S. Adams, B. Olmos, arXiv:1703.03351 (2017).}.


We thank D. Wild, E. Shahmoon, A. High, T. Andersen, N. Yao, J. Taylor, D. Greif, S. Choi, A. Keesling, R. Evans, A. Sipahigil, P. K\'omar and M. Kan\'asz-Nagy for illuminating discussions. We acknowledge funding from the MIT-Harvard CUA, NSF, AFOSR and MURI. JP acknowledges support from the Hungary Initiative Foundation. JB acknowledges funding from the Carlsberg Foundation. DEC acknowledges support from the MINECO ``Severo Ochoa'' Program (SEV-2015-0522), Fundacio Privada Cellex, CERCA Programme / Generalitat de Catalunya, and ERC Starting Grant FOQAL. Work at Innsbruck is supported by SFB FOQUS of the Austrian Science Fund, and ERC Synergy Grant UQUAM.


%

\pagebreak
\onecolumngrid
\begin{center}
\textbf{\large Supplemental Material: Topological Quantum Optics\\  in Two-Dimensional Atomic Arrays}
\end{center}
\setcounter{equation}{0}
\setcounter{figure}{0}
\setcounter{table}{0}
\setcounter{page}{1}
\makeatletter
\renewcommand{\theequation}{S\arabic{equation}}
\renewcommand{\thefigure}{S\arabic{figure}}
\renewcommand{\bibnumfmt}[1]{[S#1]}
\renewcommand{\citenumfont}[1]{S#1}

\vspace{.5cm}

\twocolumngrid

The Supplemental Material is organized as follows. In sections 1 and 2, we discuss the theory behind our calculations for the atomic lattice. In section 3, we discuss energy bands inside the light cone. In section 4, we discuss edge modes on the zig-zag boundary. In section 5, we discuss the influence of lattice size on the decay rate of edge modes. In section 6, we discuss the polarization independence of unidirectional emission. In section 7, we discuss atom-photon bound modes in the bulk of the lattice. Finally, in section 8, we discuss the effects of atomic fluctuations on the spectrum.   

\subsection{1. Dyadic Green's function in free space} 
The dyadic Green's function $G_{\alpha\beta}(\mbf r)$ in Eq.~(1) of the Main Text is the solution of the dyadic equation \cite{Dung1998_SM}
\be
\fr{\omega^2}{c^2}G_{\nu\beta}(\br)-\left(\pa_\alpha\pa_\nu-\delta_{\alpha\nu}\pa_\eta\pa_\eta\right)G_{\alpha\beta}(\br)=\delta_{\nu\beta}\delta(\br).
\ee
The Cartesian components of the Green's function are given by \cite{Dung1998_SM,Morice1995_SM}
\bal
&&G_{\alpha\beta}(\br)=-\fr{e^{ikr}}{4\pi r}\bigg[\bigg( 1+\fr{i}{kr}-\fr{1}{(kr)^2} \bigg)\delta_{\alpha\beta}\qquad\nonumber\\
&&\qquad+\bigg(-1-\fr{3i}{kr}+\fr{3}{(kr)^2}\bigg) \fr{x_\alpha x_\beta}{r^2}  \bigg]+\fr{\delta_{\alpha\beta}\delta^{(3)}(\br)}{3k^2},\qquad
\eal
where $k=\omega/c$ and $\alpha,\beta = x,y,z$ and $r=\sqrt{x^2+y^2+z^2}$.

\subsection{2. Band structure calculation} 
The modes of the periodic lattice with Bloch quasi-momentum $\bk_B$ can be obtained by substituting Eqs.~(1) and (2) from the Main Text into ${H\ket{\psi}=\hbar E_{\bk_B}\ket{\psi}}$. After transforming to a Cartesian basis using the relation ${\ket{\sigma_\pm}=\mp (\ket{x}\pm i\ket{y})/\sqrt{2}}$, finding the Bloch eigenmodes reduces to the diagonalization of the following 4x4 complex matrix 
\bal\label{eigenmatrix}
\mbf M_{\alpha\mu,\beta\nu} &=&\left(\omega_A-i\Gamma_0/2\right)\delta_{\alpha\beta}\delta_{1\mu}\delta_{1\nu} + \xi_{\alpha\mu,\beta\nu} \nonumber\\
&+&\left(\omega_A-i\Gamma_0/2\right)\delta_{\alpha\beta}\delta_{2\mu}\delta_{2\nu}+\chi_{\alpha\mu,\beta\nu},
\eal
where
\bal
\xi_{\alpha\mu,\beta\nu} = - i\mu B(\delta_{\alpha x}\delta_{\beta y}-\delta_{\alpha y}\delta_{\beta x})(\delta_{1\mu}\delta_{1\nu}+ \delta_{2\mu}\delta_{2\nu}),\qquad
\eal
gives the Zeeman splitting of the atoms and the terms accounting for the atom-atom interactions are given by 
\bal\label{chiNonBravais}
\chi_{\alpha\mu,\beta\nu} = \fr{3\pi \Gamma_0c}{\omega_A} &\Bigg[&\sum\limits_{\bR\neq 0} e^{i\bk_\text{B} \cdot \bR}G_{\alpha\beta}(\bR)\delta_{1\mu}\delta_{1\nu}\quad\nonumber\\
&+&\sum\limits_{\bR} e^{i\bk_\text{B} \cdot \bR}G_{\alpha\beta}(\bR+\mbf b)\delta_{1\mu}\delta_{2\nu}\nonumber\\
&+&\sum\limits_{\bR\neq 0} e^{i\bk_\text{B} \cdot \bR}G_{\alpha\beta}(\bR)\delta_{2\mu}\delta_{2\nu}\nonumber\\
&+&\sum\limits_{\bR} e^{i\bk_\text{B} \cdot \bR}G_{\alpha\beta}(\bR-\mbf b)\delta_{2\mu}\delta_{1\nu}\Bigg],\quad\;\;\;
\eal
where $\mbf b$ is the basis vector pointing from one site to the other within the unit cell of the non-Bravais honeycomb lattice which has two sites, $\alpha,\beta = x,y$ label the polarization components and $\mu,\nu= 1,2$ are the sublattice indices. Diagonalizing $\mbf M_{\alpha\mu,\beta\nu}$ gives four eigenvalues of the form $E_{\bk_B}=\omega_{\bk_B}-i\gamma_{\bk_B}$ for each $\bk_B$ within the Brillouin zone, where the real and imaginary parts of the eigenvalues correspond to the energies and decay rates of the modes respectively. 

To ensure rapid convergence, it is convenient to perform the summation in momentum space. We use Possion's summation formula  \cite{Antezza2009_SM,Antezza2009a_SM,Lukin2016_SM}  to obtain
\bal\label{Transformation}
&&\sum\limits_{\mathbf R\neq 0}e^{i\mathbf k_B\cdot \mathbf R}G_{\alpha\beta}(\mathbf R)= \sum\limits_{\mathbf R}e^{i\mathbf k_B\cdot \mathbf R_n}G_{\alpha\beta}(\mathbf R)-G_{\alpha\beta}(\mbf 0)\nonumber\\ 
&&\qquad\qquad\qquad=\fr{1}{\mathcal{A}}\sum\limits_{\bG} g_{\alpha\beta}(\bG-\bk_B;0)-G_{\alpha\beta}(\mbf 0),
\eal
and
\bal\label{transformationNonBravais}
&&\sum\limits_{\bR} e^{i\bk_\text{B} \cdot \bR}G_{\alpha\beta}(\bR\pm\mbf b) \nonumber\\
&&\qquad\qquad= \fr{1}{\mathcal A}\sum\limits_{\mbf G}g_{\alpha\beta}(\mbf G-\bk_\text{B}; 0)e^{\pm i\mbf b\cdot(\mbf G-\bk_\text{B})},
\eal
where $\mathcal{A}$ is the area of the periodic unit cell and the summation is performed over the reciprocal lattice vectors $\{\bG\}$ in the 2D plane, which obey $\bR\cdot \bG=2\pi m$ for integer $m$ and  $g_{\alpha\beta}(\bq;z)$ stands for the Weyl decomposition of the Green's function in terms of 2D plane waves at position $z$ along the $z$-axis. In the plane of the atoms ($z=0$) it is given by \cite{Chew1995_SM,Lukin2016_SM}
\be
g_{\alpha,\beta}(\bq;0)=\int\fr{dq_z}{2\pi}\fr{1}{k^2}\fr{k^2\delta_{\alpha\beta}-q_\alpha q_\beta}{k^2-q^2-q_z^2+i\epsilon},
\ee 
where $\bq = q_x\hat x+q_y\hat y$ and $q=|\bq|$ and  we restrict ourselves to $\alpha,\beta =x,y$. 

\begin{figure}
\centering
\includegraphics[width=0.45 \textwidth]{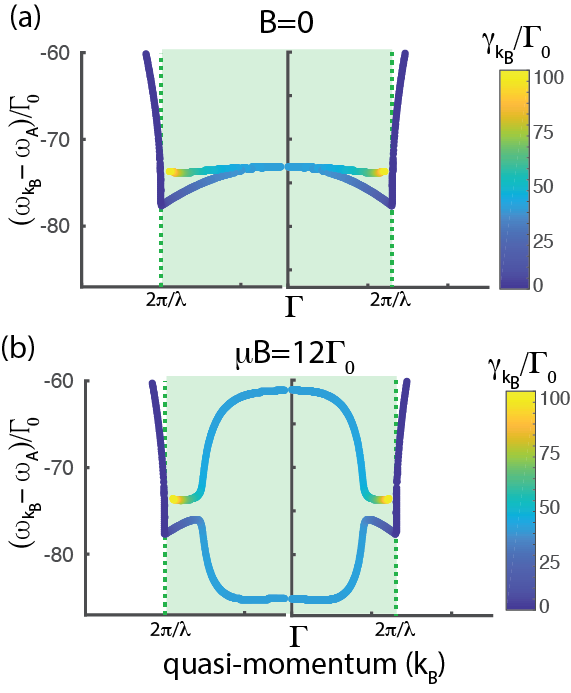}
\caption{
\label{fig:closeup} Close-up view of the lower part ($\omega_{\bk_B}<\omega_A$) of the light cone region in Fig.~1 of the Main Text. (a) When $B=0$ the two bands touch at the $\mbf \Gamma$ point, where a quadratic degeneracy is formed. The decay rate of the upper band diverges as it approaches the edges of the light cone, whereas the lower band is continuous across the edges of the light cone. (b) When a magnetic field is switched on the degeneracy is lifted.}  
\end{figure} 

The terms on the right-hand side of \eqref{Transformation} are divergent, but after regularizing both terms, their difference becomes finite. Regularization is performed by inserting a Gaussian momentum cut-off $e^{-a_\text{ho}^2(q_x^2+q_y^2+q_z^2)/2}$ into the Fourier decomposition of each term  \cite{Antezza2009_SM,Antezza2009a_SM,Lukin2016_SM}. Upon substitution, the regularized Weyl decomposition becomes 
\be
g^*_{\alpha,\beta}(\bq;0)=\int\fr{dq_z}{2\pi}\fr{1}{k^2}\fr{k^2\delta_{\alpha\beta}-q_\alpha q_\beta}{k^2-q^2-q_z^2+i\epsilon}e^{-a_\text{ho}^2(q^2+q_z^2)/2}.
\ee  
This integral can be evaluated in a closed form \cite{Lukin2016_SM} and the resulting components are given by
\bal\label{g_star}
g^*_{xx}(\bq;0) &=& (k^2-q_x^2)\mathcal{ I}(\bq),\nonumber\\
g^*_{yy}(\bq;0) &=& (k^2-q_y^2)\mathcal{ I}(\bq),\nonumber\\
g^*_{xy}(\bq;0) &=& g^*_{yx}(\bq;0) = -q_xq_y\mathcal{ I}(\bq),
\eal
where we have defined
\bal
\mathcal I(\bq) = \chi(\bq)\fr{\pi}{\Lambda(\bq)}[-i+\text{erfi}(a_\text{ho}\Lambda(\bq)/\sqrt{2})]
\eal
with
\bal
\chi(\bq)&=&\fr{1}{2\pi k^2}e^{-a_\text{ho}^2(q_x^2+q_y^2+\Lambda(\bq)^2)/2}
\eal
and
\bal
\Lambda(\bq) &=& (k^2-q_x^2-q_y^2)^{1/2},
\eal
where $\text{Im}(\Lambda)\geq 0$ and $\text{Re}(\Lambda)\geq 0$ is assumed and $\text{erfi}(x)$ stands for the imaginary error function. 
The regularized Green's function at the source takes the form \cite{Antezza2009_SM} 
\bal\label{G_reg}
G^*_{\alpha\beta}(\mbf 0)&=&\fr{k}{6\pi}\bigg[\bigg(\fr{\text{erfi}(ka_\text{ho}/\sqrt{2})-i}{e^{(ka_\text{ho})^2/2}}\bigg)\nonumber \\
&&\qquad\quad- \fr{(-1/2)+(ka_\text{ho})^2}{(\pi/2)^{1/2} (ka_\text{ho})^3}\bigg]\delta_{\alpha\beta}.
\eal
It can be shown using the methods developed in \cite{Antezza2009_SM} that $\exp(k^2a_\text{ho}^2/2)[1/\mathcal{A}\sum_{\bG} g^*_{\alpha\beta}(\bG-\bk_B;0)-G^*_{\alpha\beta}(\mathbf 0)]$ approaches the value of the left-hand side of \refeq{Transformation} as the limit $a_\text{ho}\to 0$ is taken \cite{Lukin2016_SM}. Therefore, choosing a small $a_\text{ho}$ such that $a_\text{ho}\ll \lambda$ we obtain
\bal\label{TransformationRegularized}
&&\sum\limits_{\mathbf R\neq 0}e^{i\mathbf k_B\cdot \mathbf R}G_{\alpha\beta}(\mathbf R)\nonumber\\ 
&&\qquad\approx\fr{e^{k^2a_\text{ho}^2/2}}{\mathcal{A}}\sum\limits_{\bG} g^*_{\alpha\beta}(\bG-\bk_B;0)-G^*_{\alpha\beta}(\mbf 0),\quad
\eal
where $g^*_{\alpha\beta}$ is given by \refeq{g_star} and $G^*_{\alpha\beta}$ is given by \refeq{G_reg}. The summation over $\mbf G$ converges rapidly.

\subsection{3. Energy bands within the light cone}


\reffig{fig:closeup} shows a close-up view of the lower part (${\omega_{\bk_B} < \omega_A}$) of the light cone region of Fig.~1 of the Main Text. For $B=0$, the two bands are close to each other in energy and touch at the $\mbf \Gamma$ point in a quadratic degeneracy. The quadratic degeneracy arises due to the inversion symmetry of the reciprocal lattice with respect to the $\mbf \Gamma$ point \cite{Chong2008_SM}. Switching on a magnetic field raises the degeneracy and the bands exchange one unit of Berry flux ($\Delta C_\pm =\pm 1$). The modes in the lower band are predominantly polarized in the longitudinal direction, whereas the modes in the upper band have transversal polarization. Therefore, the lower band couples weakly to the transversely polarized free-space modes and remains continuous as it crosses the edges of the light cone. In contrast, the upper band couples strongly to free-space modes and the decay rate of the modes diverges as the edges of the light cone are approached, effectively `dissolving' the band due to broadening. In particular, for the upper band $\gamma \sim 1/\sqrt{1-(k_B/k_L)^2}$, where $k_B$ is the magnitude of the in-plane Bloch vector and $k_L=2\pi/\lambda$ \cite{Shahmoon2017_SM}. Physically, the divergence arises due to the fact that a free-space photon traveling exactly in-plane would interact with an infinite number of atoms and, therefore, a transversely polarized extended lattice mode with $k_B=2\pi/\lambda$ would decay immediately through its overlap with free-space modes of the same momentum \setcounter{footnote}{10}\footnote{We thank D. S. Wild for pointing this out}.

\begin{figure}
\centering
\includegraphics[width=0.45 \textwidth]{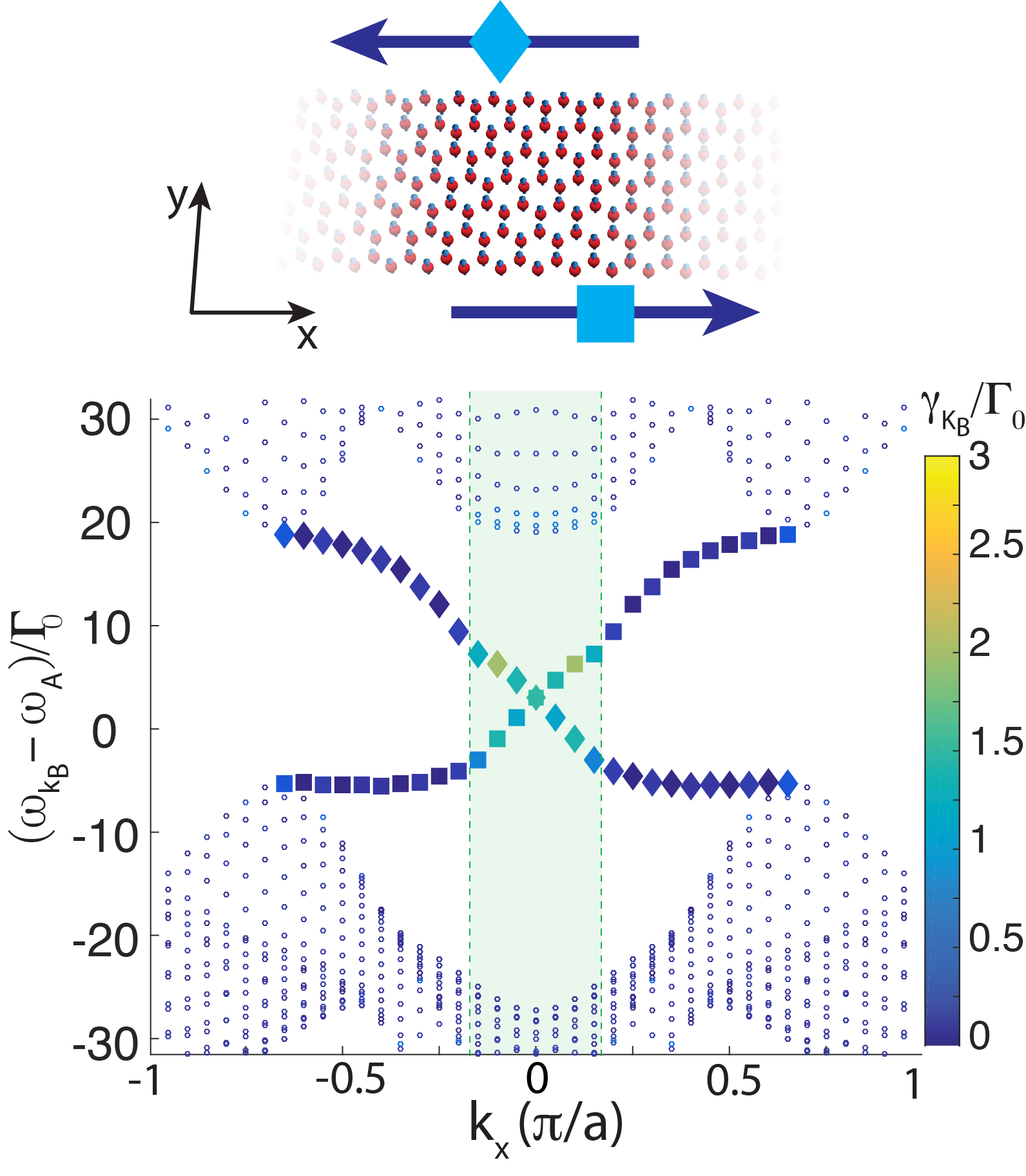}
\caption{
\label{fig:zigzag} Edge modes on the zig-zag boundary of a honeycomb lattice. Edge modes in the bottom half of the band gap have quasi-momenta $k_B<\omega_{\bk_B}/c$ and are short-lived. Modes in the top half of the gap have quasi-momenta $k_B>\omega_{\bk_B}/c$ and are long-lived. Relevant parameters are $\lambda=790\text{nm}$, ${\Gamma_0=2\pi\times6\text{MHz}}$, $a=0.05\lambda$ and $\mu B = 12\Gamma_0$ and the spectrum was obtained from a lattice with 40x41 atoms with periodic boundary conditions along the first dimension. States for which the ratio of the total amplitude on the top (bottom) four atom rows to the bottom (top) four rows is greater than 15 are classified as edge states.}  
\end{figure}

\subsection{4. Edge modes on the zig-zag boundary} In addition to the bearded and armchair modes, a honeycomb lattice can also be terminated by a zig-zag boundary. Fig.~\ref{fig:zigzag} shows the spectrum of the edge modes on such a boundary. Edge modes in the bottom half of the band gap have quasi-momenta $k_B<\omega_{\bk_B}/c$ and are short-lived, whereas modes in the top half of the gap have quasi-momenta $k_B>\omega_{\bk_B}/c$ and are long-lived. By tuning the frequency of a laser to be resonant with the modes in the top half of the gap, we can predominantly excite the long-lived edge modes of the zig-zag edge.

\begin{figure}
\centering
\includegraphics[width=0.35 \textwidth]{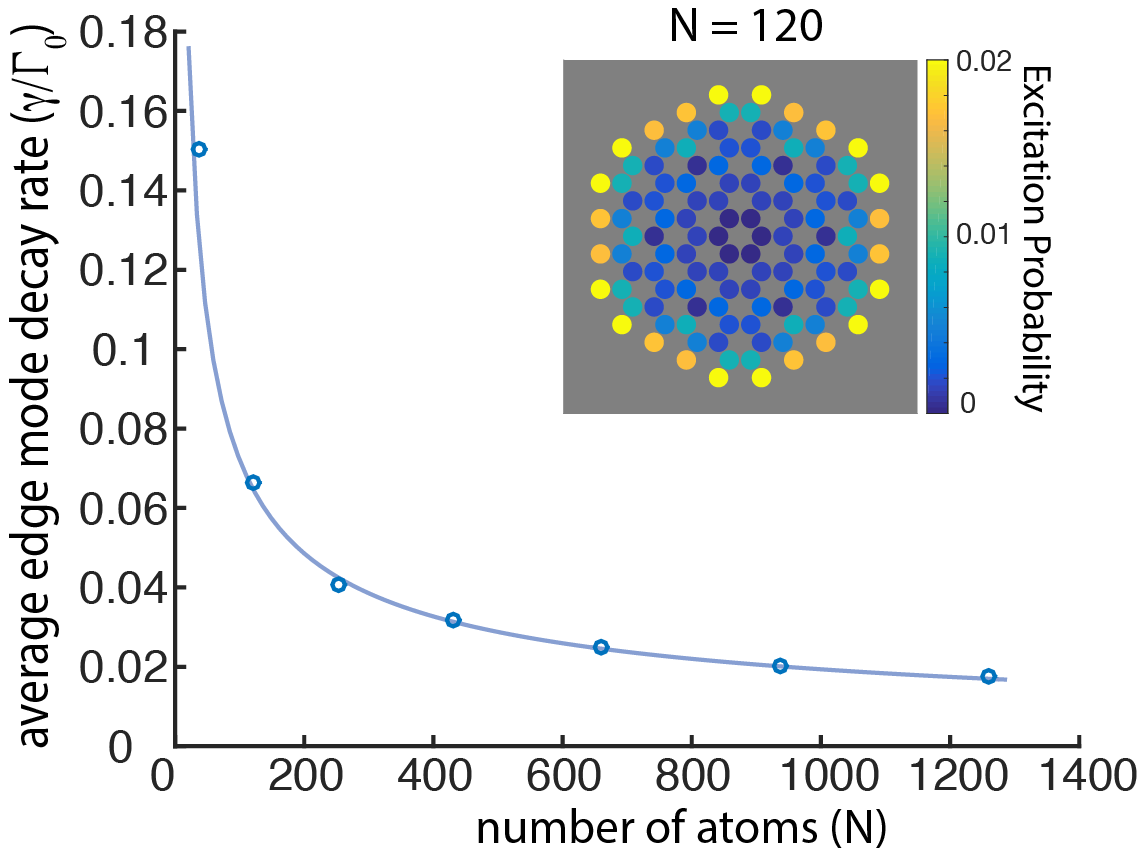}
\caption{
\label{fig:atomno} Scaling of the decay rate of the edge modes on the bearded boundary as a function of the total number of atoms in the hexagonal-shaped atom lattice. The decay rate scales approximately as $\sim 1/\sqrt{N}$. The numerical fit to the data points $\sim1/N^{0.57}$ is shown as a solid line. To obtain each point in the plot, a hexagonal-shaped lattice of $N$ atoms with bearded boundaries was considered with $a=\lambda/20$ and $\mu B = 12\Gamma_0$. The average decay rate was obtained by averaging the decay rates of all edge states inside the band gap. The inset shows the probability amplitudes of a bearded edge state on a lattice of $N=120$ atoms.  } 
\end{figure}  

\subsection{5.  Influence of system size on decay rate of edge modes}
For an infinite lattice, modes with quasi-momenta $k_B>\omega_{\bk_B}/c$ are decoupled from free-space modes and, therefore, do not decay. In contrast, for finite lattices even such modes have a finite lifetime. Fig.~\ref{fig:atomno} shows the decay rate of the bearded edge modes with $k_B>\omega_{\bk_B}/c$ as the total number of atoms in the lattice is varied. The interatomic spacing is assumed to be fixed at $a=\lambda/20$. The decay rate of the modes scales approximately as $\sim 1/\sqrt{N}$, where $N$ is the total number of atoms in the 2D lattice. This scaling is consistent with the observation that, since edge modes are confined to the boundaries, they explore only the 1D perimeter of the lattice, which scales with $\sim\sqrt{N}$.

Note that changing the size of the lattice has little effect on the decay rate of short-lived edge modes with quasi-momenta $k_\text{B} < \omega_{\bk_B}/c$. These modes couple directly to free-space modes and thus their loss via out-of-plane emission dominates, making finite-size effects negligible in comparison. 

\subsection{6.  Polarization independence of unidirectional emission }

In Fig.~(4) of the Main Text a single atom on the lattice boundary is driven by a laser. When the transitions to the $\ket{\sigma_+}$ and $\ket{\sigma_-}$ states are driven with equal coupling strengths $\Omega$, approximately 96\% of the excitation emitted by the driven atom is coupled in the edge modes carrying energy in the forward direction.   
If only one of the transitions to the $\ket{\sigma_-}$ or $\ket{\sigma_+}$ states is driven, the efficiency of coupling into the unidirectional edge modes changes to approximately 90\% and 97\%, respectively. 

Note that the fact that unidirectional emission does not depend on which transition of the atom is driven demonstrates that the unidirectionality arises from topology and not from polarization selection as, for example, in Ref.~\cite{Mitsch2014_SM}.

\begin{figure}
\centering
\includegraphics[width=0.4 \textwidth]{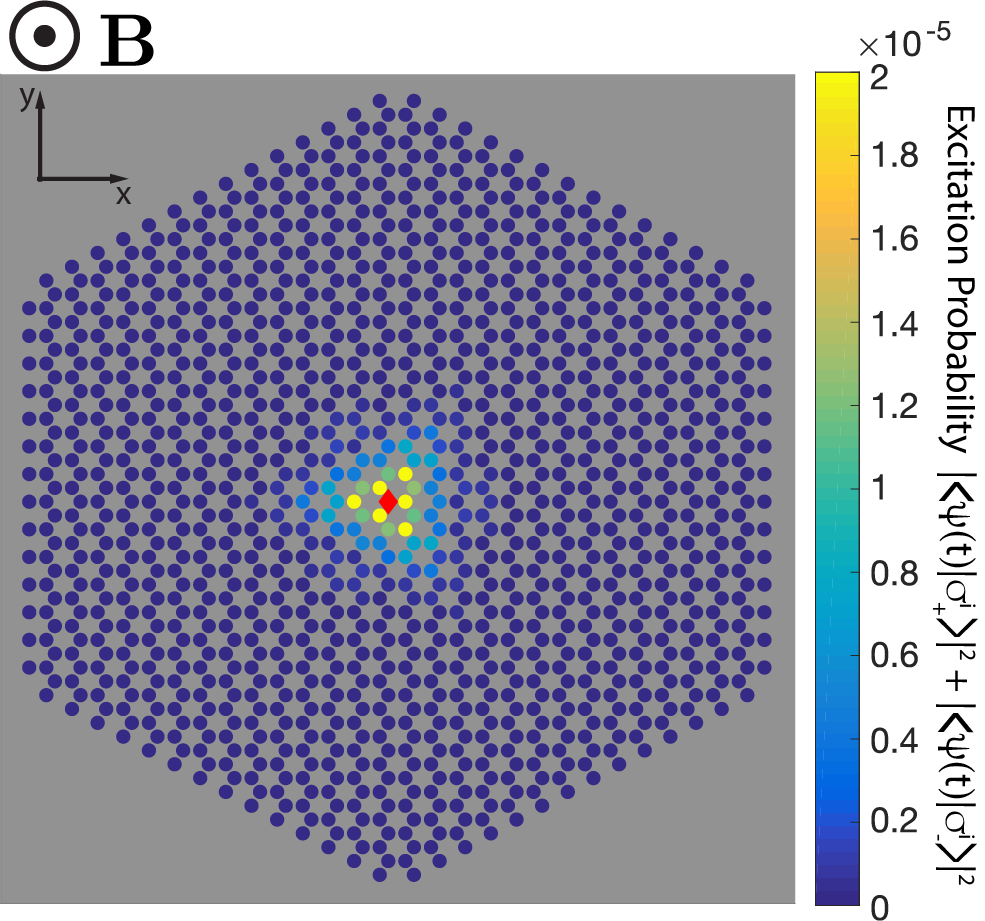}
\caption{
\label{fig:bulkexcitation} A sub-radiant atom-photon bound state forms when an atom in the bulk of the lattice (marked with a red diamond) is driven by a laser with its frequency inside the band gap. The decay rate of the bound state is sensitive to the polarization of the driving laser. Relevant parameters are $N=1260$, $\lambda=790\text{nm}$, ${\Gamma_0=2\pi\times 6\text{MHz}}$, $a=0.05\lambda$ and $\mu B = 12\Gamma_0$. The strength of the drive is $\Omega = 1\Gamma_0$ and the frequency of the laser is tuned such that $\omega_\text{L}=\omega_A + 10\Gamma_0$. The driving laser is adiabatically switched on with a sigmoid profile $\Omega(t) = \Omega\left[1+\exp\left(-\left(t-3\Gamma_0^{-1}\right)/0.3\Gamma_0^{-1}\right)\right]^{-1}$.   The figure shows a snapshot at $t=10\Gamma_0^{-1}$.} 
\end{figure} 

\subsection{7.  Bulk excitations within the band gap}

In the Main Text we discussed the system dynamics when edge states are excited through an individual atom that is located near the boundary of the lattice. Here we focus on the time evolution of the system when an atom in the bulk is excited with a laser, whose frequency $\omega_\text{L}$ falls inside the band gap. The driving laser is adiabatically switched using a sigmoid profile to avoid exciting non-resonant modes and it continuously excites the atom.   

Since inside the band gap there are no extended bulk modes, the atom cannot resonantly couple to any of the extended lattice modes. Instead, the atom weakly dresses the far-detuned modes. Consequently, the atom exchanges energy only with atoms in its immediate neighborhood and a spatially confined atom-photon bound state is formed as shown in \reffig{fig:bulkexcitation}. This is analogous to the atom-photon bound states that are predicted to exist in photonic crystals with band gaps \cite{John1990_SM,Douglas2015_SM}.  

Here a sigmoid profile is preferred to a Gaussian one, since the higher order derivatives of the sigmoid function vary slower as the function approaches its maximum value than the corresponding derivatives of a Gaussian profile. Thus the sigmoid profile performs better than a Gaussian in not exciting far detuned extended bulk modes, making the weak, off-resonant dressing of bulk modes observable. 

Since the majority of the extended bulk modes above and below the band gap are long-lived, the bound state itself is sub-radiant with a decay rate that depends on the polarization of the exciting laser. In particular, since the band gap arises from the Zeeman-splitting of the $\ket{\sigma_+}$ and $\ket{\sigma_-}$ levels, the bulk modes above and below the band gap couple more strongly to light polarized along $\hat \sigma_+$ and $\hat \sigma_-$ respectively. Since the modes close to the center of the Brillouin zone and immediately above the gap are short-lived, a laser with polarization $\hat\sigma_+$ excites a shorter lived bound state  with $\gamma=\Gamma_0/4.7$, whereas a laser with polarization $\hat\sigma_-$ yields $\gamma=\Gamma_0/7.7$. For an $\hat x$ polarized laser we obtain $\gamma = \Gamma_0/5.7$. 
 \begin{figure}[h!]
\begin{center}
\includegraphics[width=0.4 \textwidth]{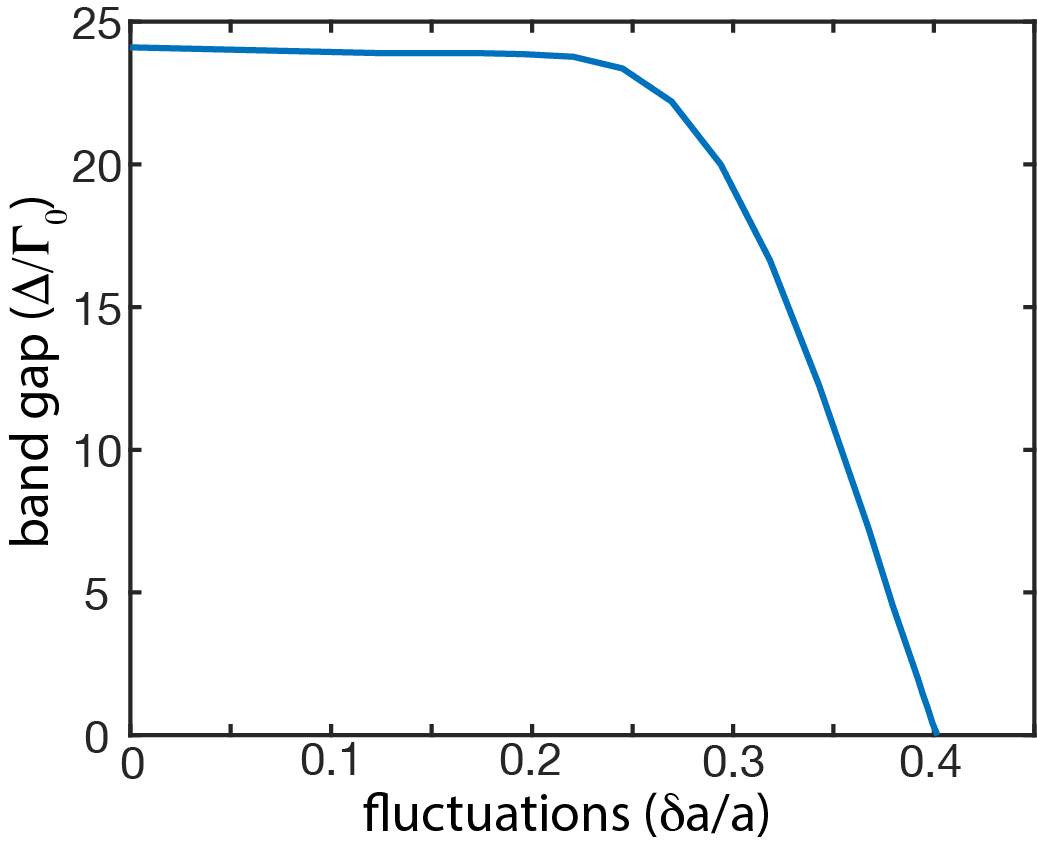}
\caption{
\label{fig:fluctuations} Maximum gap size $\Delta_\text{max}$ as the atomic positions fluctuate with amplitude $\delta a$ around the lattice sites with fixed interatomic spacing $a=0.05\lambda$ and $\mu B = 12\Gamma_0$. When the amplitude of the fluctuations is less than 25\% of the interatomic spacing $a$, the band structure is not significantly affected.} 
\end{center}
\end{figure} 
 
\subsection{8. Effect of fluctuating atomic positions} 
The Hamiltonian in Eq.~(1) of the Main Text assumes that the position of the atoms is fixed at the sites of the lattice. In practice, even when the atoms are tightly trapped and are occupying their motional ground state, their position will fluctuate around the lattice sites. These quantum fluctuations are uncorrelated between different sites. To quantify how the quantum fluctuations in atomic positions affect our results, we assume a harmonic trapping potential of frequency $\omega_\text{ho}$, with the corresponding spatial extent of the ground state oscillations on the order of $\delta a=\sqrt{\hbar/(2m\omega_\text{ho})}$. We then average the dyadic Green's function in the Hamiltonian with respect to the ground state fluctuations \cite{Antezza2009_SM,Lukin2016_SM}. \reffig{fig:fluctuations} shows how the size of the band gap between the topological bands changes as the magnitude of $\delta a$ is varied as a fraction of the interatomic spacing $a$. Larger fluctuations smear out the well-defined phase between different atoms and eventually the gap closes. However, when the extent of the fluctuations is less than $25\%$ of the interatomic spacing, the size of the gap (and the band structure as a whole) is not significantly affected. This shows that our results are robust against moderate fluctuations in atomic positions around the lattice sites.

\end{document}